\documentclass[conference]{IEEEtran}
\usepackage[tight,footnotesize]{subfigure}
\usepackage{array}
\usepackage{makeidx}
\usepackage[hyphens]{url}
\usepackage{hyperref}
\usepackage[gennarrow]{eurosym}
\usepackage{wasysym}
\usepackage{graphicx}

\makeatletter
\usepackage{url}
\usepackage{csquotes}
\date{}

\makeatother

\begin{document}

\title{The Nuts and Bolts of Micropayments: a Survey}

\author{\IEEEauthorblockN{Syed Taha Ali}
\IEEEauthorblockA{NUST School of Electrical Engineering\\ and Computer Science, Pakistan\\
Email: taha.ali@seecs.edu.pk}
\and
\IEEEauthorblockN{Dylan Clarke}
\IEEEauthorblockA{Newcastle University\\
United Kingdom\\
Email: dylan.clarke@newcastle.ac.uk}
\and
\IEEEauthorblockN{Patrick McCorry}
\IEEEauthorblockA{University College London\\
United Kingdom\\
Email: stonecoldpat@gmail.com}}

\maketitle
\begin{abstract}

We are witnessing a veritable explosion of interest in new electronic payments systems and modalities, such as digital wallets, mobile and contactless payments, and cryptocurrencies such as Bitcoin. One area of research and commercial interest at the confluence of these trends, which is also receiving reinvigorated attention, is micropayments. Indeed, a workable micropayments system, one that lets users purchase digital content in an easy and ``hassle-free'' manner with payments in the order of cents and lower, has long been regarded as the holy grail of web-publishing. The research community has actively worked on this problem over the past two decades, numerous creative solutions have been presented, business ventures have been launched, but a mainstream solution has yet to emerge.

In this paper, we undertake a comprehensive survey of key trends and innovations in the development of research-based and commercial micropayment systems. Based on our study, we argue that past solutions have largely failed because research has focused heavily on cryptographic and engineering innovation, whereas fundamental issues pertaining to usability, psychology, and economics have been neglected. We contextualize the range of existing challenges for micropayments systems, discuss potential deployment strategies, and identify critical stumbling blocks, some of which we believe researchers and developers have yet to fully recognize. We hope this effort will motivate and guide the development of micropayments systems.

\end{abstract}


\section{Introduction}

The ongoing popularity of Bitcoin has inspired keen interest in digital currencies in the research community, the financial sector, and even at the government level. This surge has also rekindled the conversation on developing systems to enable micropayments, i.e. low-value digital transactions, typically in the order of pennies and cents. Micropayment transactions may be considered the electronic equivalent of purchases made using pocket cash or spare change. Historically, the problem with low-value transactions has been that processing and transaction fees end up dwarfing the actual transaction amount\footnote{For instance, in 2014 UK-issued debit and credit cards (with chip and pin) typically averaged transaction fees of 14 and 81.5 pence respectively \cite{hart2014breakdown}.}. Payment processors impose these fees for a variety of reasons including infrastructure costs, administrative charges, and mechanisms for fraud prevention and dispute resolution. There has been considerable research in the past two decades on using digital communications and cryptography to minimize these costs, ideally down to the fraction-of-a-cent range.

The traditional argument goes that, if enabled, micropayments stand to be a key pillar of the information economy \cite{schmidt1999framework}, with direct and immediate applications in reviving journalism \cite{fisher2009saving} and supporting the music industry \cite{sisario2013music}. The ability to economically transfer minuscule amounts of money at high speeds will empower dynamic new pricing models where digital content such as online newspapers, magazines, and music albums can be unbundled, allowing consumers to purchase individual news stories, articles, and songs. Furthermore, with pricing in the sub-dollar range, users will be encouraged to increase spending and also engage in impulse purchases, thereby opening up powerful new revenue streams.

There have been two main waves of innovation in designing and deploying micropayment systems, the first in the late 1990s and the second in the 2000s \cite{lesk2004micropayments} \cite{parhonyi2011micropayment}. Both efforts largely failed, and only a few systems have survived. Reasons include poor infrastructure support, cumbersome and non-intuitive system design, and conservatism on the part of financial institutions and users. Critics have also cited poor business cases and neglect of critical psychological factors \cite{szabo1999micropayments} \cite{shirky2000case} \cite{odlyzko2003case}.

Today, however, the landscape has changed in some fundamental ways. First and foremost, the business case for micropayments is validated. Large numbers of consumers now regularly make low-value payments for online content. Apple's iTunes store has proved a resounding success \cite{griggs2013apple}. In the smartphone universe, iOS app developers reportedly made over \$10 billion in 2014 from in-app purchases, to put in context, a figure greater than Hollywood's box office earnings \cite{price2015app}. New multiplayer video games now enable millions of players to make in-game purchases as part of gameplay. The popular League of Legends singlehandedly earned \$624 million in 2013, and almost a billion dollars in 2014 \cite{chalk2014league}, from these `microtransactions' in which players purchase premium in-game items like characters, weapons, healing portions, etc. costing single digit dollar amounts.

Second, the current advertising-based web publishing model is in crisis. Web ads are intrusive, degrade user experience, and significantly increase data consumption \cite{roberts2015ad}, a particular concern for mobile users \cite{salmon2015ad}. Users are also concerned about privacy and third party tracking \cite{chanchary2015user} \cite{agarwal2013not} \cite{ur2012smart}, especially in the wake of the Snowden revelations \cite{annalect2013online}.

Collectively these factors have given rise to the `ad-wars' phenomenon: globally, some 198 million people deploy adblocking software, such as Ghostery and Adblock Plus, leading to a staggering \$22 billion in lost revenues \cite{pagefair2015cost}. Upcoming versions of Apples IoS9 and OSX 10.11 are both reported to feature default ad blocking functionality \cite{naughton2015this}. A recent study \cite{pagefair2015cost} notes that adblockers pose an ``existential threat" to the ad-based publishing model. Some commentators are therefore calling for a fundamental rethink of the current web publishing model  \cite{stern2015from} \cite{isaacson2014big}. Micropayments are a leading alternative.

Third, there are promising developments on the ground. The technology has vastly improved in the last decade: high speed broadband is ubiquitous, public key infrastructure is widely deployed, Web browsers have far more functionality, and smartphone penetration is high.  Public attitudes have also changed: millions regularly engage in online banking and participate on social networks. The concept of mobile wallets and cryptocurrencies is no longer alien. Surveys report people are now more willing to pay for online content \cite{plimmer2013britons} \cite{lu2015more}. Charities have begun to leverage micropayments (or `microdonations') for raising funds \cite{radojev2016pennies}.

There is also the Bitcoin experiment. Whereas Bitcoin's long term success is still an open question, its popularity has nonetheless inspired researchers to reimagine payments systems.  Financial institutions and governments also appear more receptive to innovation. Some of the world's largest banks are already in the process of appropriating Bitcoin's key innovation, the blockchain, to reduce infrastructure costs by an estimated \$15-20 billion \cite{khaw2015nine}.

Due to these factors we are witnessing what we believe is the third wave of micropayment systems. Several new micropayments solutions have launched, several more are about to, and collectively several millions of dollars of startup capital has been raised. Blendle, an ``iTunes for newspapers'', has received substantial press, and has made deals with the New York Times, the Washington Post, and the Wall Street Journal to sell individual articles for 20 cents on average \cite{reilly2015new}. WeChat, a leading Chinese content publishing platform, and one of the world's largest, with 600 million active users intends to introduce a `like' button which will allow readers to reward authors with micropayment donations ranging from under \$1 up to \$30 \cite{horwitz2015like}. Google Contributor allows users to pay a small monthly fee for an ad-free browsing experience on supported websites \cite{gibbs2014google}. And a slew of solutions, such as Bitwall, BitMonet and Flattr, piggyback on Bitcoin's payment network \cite{cawrey2013bitcoin} taking advantage of Bitcoin's low transaction fees\footnote{At the time of writing, the minimum transaction fees for Bitcoin Core version 0.11.1 stands at 0.00005 satoshis which equates to \$0.02 \cite{bitcoincore2015bitcoin}.}. Brave Software Inc. is currently trialling a micropayments solution integrated directly into the Brave Web browser \cite{scott2016brave}.

We believe therefore that this is an opportune time to revisit the topic of micropayments. Our contributions are:

\begin{enumerate}
\item We undertake a comprehensive survey of micropayments solutions in the research literature and highlight the workings and key features of representative systems,

\item We classify past and present commercial micropayments systems and identify the strategies they use,

\item We identify key challenges ahead in design and deployment of these systems and formulate recommendations.

\end{enumerate}

Based on our study, we find is that there is considerable room for work. Research-based systems consist almost entirely of novel cryptographic solutions with the primary design focus being security and efficiency concerns, whereas commercial systems opt for simple and intuitive cost-cutting strategies such as  aggregating multiple payments and automating payment processing. These two domains are mostly isolated from each other (with the notable exception of Bitcoin-based micropayments systems). However, the vast majority of micropayments solutions have failed, in large part due to neglect of critical non-technical concerns such as usability issues, ethical and legal concerns, poor business cases, and ineffective deployment strategies. However, we believe that once these challenges are fully recognized, technology may be successfully used to address them. For this reason, we do not restrict our study solely to the cryptographic literature, but also draw together critical insights from other domains impacting micropayments systems.

To the best of our knowledge, we are the first to perform such a broad study. We have located only two prior surveys on micropayment systems in the past decade: P{\'a}rhonyi \cite{parhonyi2011micropayment} documents micropayments systems with a focus on commercial solutions, whereas Jain et al. \cite{jain2008peer2peer} specifically consider peer-to-peer schemes from the research literature. Surveys on digital currencies (e.g. \cite{venkataiahgari2006survey} \cite{belenkiy2011cash}) usually include some micropayments schemes but the emphasis is on aspects of electronic cash in general. Micropayments systems involve certain unique challenges (technological, psychological, and economic) differentiating them from general payments systems and necessitate a specialized study.

The rest of this paper is organized as follows: in Sec.~\ref{sec:background}, we introduce key properties of micropayments systems and broadly summarize developments in this field. In Sec.~\ref{sec:cryptography_based_systems}-\ref{sec:commercial_solutions}, we examine the range of cryptographic and commercial solutions and emphasize their strengths and weaknesses. In Sec.~\ref{sec:outstanding_challenges_and_future_directions}, we discuss key challenges ahead and present recommendations. We conclude in Sec.~\ref{sec:conclusion}.

\section{Background}
\label{sec:background}

In this section, we qualify micropayments, discuss properties of micropayments systems and trace their development.

\subsection{Definitions and Properties}

There is considerable variance over how small a payment must be to qualify as a micropayment. One of the earliest solutions, Millicent, envisioned transactions in the sub-penny range \cite{manasse1995millicent}. Kniberg classifies them as payments of up to \euro1 \cite{kniberg2002makes}, whereas a study of European online payments sets the threshold at \euro 5 \cite{carat2002epayment}. Commercial provider PayPal classes micropayments as typically under \$10 \cite{paypalmicropayments}. There is, however, broad agreement that the associated processing fees should be low enough to justify very small transactions and that these costs should ideally be significantly less than those charged by mainstream payment systems, such as credit cards.

The size of the payments broadly determine the requirements of the payment system. For macropayments (i.e. large and medium-sized payments), regulation may mandate that payments be recorded and that dispute-resolution mechanisms be implemented. Customers themselves may prefer extensive transaction records and fraud prevention mechanisms for large value payments, all of which result in higher processing costs. Furthermore, users typically make large transactions much less frequently than smaller ones and processing fees for the former may not seem too heavy a burden. User anonymity and processing fees are therefore generally of secondary importance in macropayments and assume primary concern as payment size approaches the sub-dollar range.

\textbf{Processing fees} for payments systems generally comprise infrastructure and clearing costs, i.e. costs due to equipment, computation, storage, communication and accounting. Certain systems may incur additional costs, depending on transaction type or payment modality. However, micropayments also involve what economists describe as cognitive or \textbf{mental transaction costs}, i.e. the ``hassle-factor'' associated with having to choose fine-grained bundling options at very low prices. Szabo \cite{szabo1999micropayments} has argued persuasively that researchers often overlook the fact that these mental costs outweigh the technological and are a determining factor in system adoption.

In the context of e-commerce, micropayment systems are generally envisioned as incurring minimal processing delay and facilitating instant delivery of goods \cite{parhonyi2005micro}. The micropayments ecosystem typically consists of three principal entities: \textbf{customer} or user denotes an individual or party which transacts goods and services from a \textbf{merchant}. The transaction is enabled or facilitated by a \textbf{broker}. This role often belongs to banks or financial institutions which issue the financial instrument or currency used in the transaction, maintain balance accounts for customers and merchants, redeem their funds, and arbitrate in dispute resolution\footnote{We use the terms broker and bank interchangeably in this paper.}. Some systems may involve other entities such as peers, certificate authorities, or trusted third parties for various purposes.

System design and uptake is determined by which properties the system provides. We discuss here key properties pertaining to micropayments systems:

\textbf{Anonymity:} refers to the exposure of the customer's identity and personal information to the merchant and the broker as a direct result of using the payment system. Anonymity in this sense is synonymous with customer privacy. Complete anonymity is achievable with physical cash, whereas with a credit card, both merchant and bank are privy to the customer's identity and her purchase details. Some commercial systems offer strong anonymity. An example is paysafecard where customers acquire prepaid scratchcards from shops for later spending without revealing any personal information.

Considerable legal issues come into play as a truly anonymous payment mechanism can become a tool for money laundering and crime. Some systems achieve partial anonymity by using pseudonyms to shield personal information from merchants but not the banks. Indeed, in the majority of commercial systems we survey, anonymity from the bank is not a design goal. Another notion is that of revocable anonymity where customer privacy may be overturned in the event of disputes \cite{stadler1996cryptographic}. Anonymity is ensured either through employing cryptography or defining special procedures. Anonymity is typically established when the customer acquires the currency from the broker or when she pays the merchant.

\textbf{Security:} refers to the integrity of the system, its resilience to fraud, and in particular its ability to prevent counterfeiting and double-spending. In most commercial systems such as PayPal, the broker maintains a customer balance and explicitly authorizes every transaction. Bitcoin extends this approach by employing a distributed and highly synchronized ledger called the blockchain. A novel strategy, used by MicroMint, is to use cryptography to mint crypto-tokens which are far too difficult and expensive to counterfeit.

\textbf{Validation:} indicates whether a system requires real-time contact with the broker to process transactions. Payments systems in the 1990s were limited by low-speed and unreliable dial-up Internet access. Furthermore, requiring the broker in every transaction effectively rendered him a communications bottleneck and a single point of failure in the system. Some solutions, referred to as `optimistic' \cite{asokan1996optimistic}, resolved this issue by contacting the broker only for a small subset of transactions, generally those that proved exceptional or problematic. However, this restriction is now considerably relaxed due to ubiquitous high-speed broadband access and the prevalence of cloud computing.

\textbf{Transferability:} denotes the ease and extent of transferring funds using the system. This includes notions of system \textbf{coverage, acceptability,} and \textbf{penetration} among customers and merchants. Some systems permit a wider range of transactions, such as peer-to-peer transactions in which users may transfer funds directly to each other. Transferability also includes \textbf{interoperability}, i.e. permitting payments between different systems and financial institutions, and \textbf{versatility}, i.e. facilitating different payment options, such as offline payments, payments using handheld devices, etc.

\textbf{Payment Mode:} indicates how a system actually undertakes the transfer of value between parties. Pre-paid (or debit-based) systems require the customer to input funds into the system prior to making payments which are later deducted from her account. Post-paid (or credit-based systems) track customer spending and charge her at the end of the billing period.

Certain properties apply only to select systems. For example, \textbf{divisibility}, i.e. the ability of the system to make payments of arbitrary value, is a limiting factor for some token-based systems. Some systems assume specific \textbf{relationship models}, i.e. the service or user experience may be different depending on whether the customer and merchant have a long-term or persistent business relationship as opposed to casual or transient interactions. Some systems may be \textbf{hardware-reliant}, i.e. the payment solution relies on a physical device or card which stores cryptographic credentials or currency units. This category includes mobile wallets, smartcards such as Octopus, and Bitcoin wallets like Trezor \cite{trezor} and Case \cite{case}.

In conclusion, there are certain properties common to electronic and networked systems in general. These include \textbf{usability}, the ease of use of a system, \textbf{scalability}, the ability of a system to handle increasing numbers of users and larger payment volumes without significantly degrading performance, and \textbf{reliability}, the measure of how dependable a system is.

Next we briefly summarize developments in this field.

\subsection{A Short History of Micropayment Systems}

The motivation for micropayments derives primarily from the notion of the \textbf{information economy} \cite{schmidt1999framework}. Digital goods such as music mp3 files, blogposts, and software are distinct from their physical counterparts, CDs, newspapers, etc. in fundamental ways. For one, digital goods typically bear high fixed costs but negligible marginal costs, i.e. they are expensive to produce but the cost of reproducing these goods is near zero. Deriving from this notion, digital goods are also \emph{non-rival goods}, i.e. they are not restricted to a single customer and may be consumed by multiple parties simultaneously.

These characteristics also apply to digital services which include not just traditional services such as stock quotes and newspapers delivered in electronic format but also interactive new paradigms such as massive multi-player online games (MMOGs). Furthermore, the Internet enables mass distribution of digital goods and services at low-cost: not only do customers have greater access to goods and services but electronic transaction costs are also significantly smaller. In this scenario, affixing very low fees to digital goods and services can prove a powerful source of long-term revenue for merchants.

Micropayment systems for digital content were envisioned as early as the 1960s, when visionary Ted Nelson conceived of intrinsically bidirectional hyperlinks, enabling users to electronically pay for content that they access \cite{nelsonathought}. In the late 1990s, pioneers Tim Berners-Lee and Marc Andreessen considered incorporating micropayments directly into the Web at the protocol level but were discouraged by conservative banking regulations \cite{isaacson2014bitcoin}.

Observers broadly agree that there have been two waves of innovation in micropayment systems \cite{lesk2004micropayments} \cite{parhonyi2011micropayment}.

The first generation of systems surfaced in the mid-1990s, inspired by the electronic cash movement led by DigiCash. In keeping with the dial-up Internet infrastructure of the time and the relatively low processing power on computers, the main design goal for these systems was to minimize communication and computation costs. A popular strategy used by by these systems was the \textbf{account-based approach}, used by systems such as CyberCoin, Mini-Pay and NetBill.  In this case, the broker maintained accounting ledgers for customers and merchants and aggregated net flow of funds in and out of the system. A centralized ledger prevents double-spending and fraud, whereas aggregation of multiple low-value payments effectively amortizes the processing fees incurred in moving funds between the system and the banking infrastructure. A second strategy, the \textbf{crypto-token approach}, employed by systems such as PayWord and MicroMint, used cryptographic mechanisms to mint digital tokens which were computationally infeasible to counterfeit. Customers then used these tokens to transact with merchants.


Several of these solutions were commercialized in partnership with well-known brands, such as DEC, IBM, and Visa but efforts were to fail for a variety reasons. The dial-up infrastructure was slow and unreliable. These systems had poor usability and suffered from high latency. CyberCash, for example, typically took 15-20 seconds to finalize a transaction \cite{weber1998chablis}. Systems such as Mondex and CAFE also required trusted hardware. Furthermore, the business models have been heavily criticized \cite{shirky2000case} \cite{odlyzko2003case}, the systems had poor interoperability, and consequently there was low penetration among merchants.

The second generation of micropayment systems, appearing around 2000, were mostly account-based offerings, such as PayPal and ClickandBuy. Users transferred funds into the system and then used the amount to make payments. This approach had a physical counterpart in prepaid cards, such as Wallie and paysafecard, which users purchased at shops for fixed denominations and then progressively `spent' in online transactions. Another innovation was the use of communications infrastructure, such as email and mobile phones, to validate transactions, as done in systems like Zong.

Several of these systems have survived to the present day. Reasons include good usability, intuitive design, and low latency. For instance, ClickandBuy puts specific icons on merchant websites, which users click to access the payment portal, following which they get immediate access to purchased content. Legislation has evolved to protect users from fraud and safeguard their privacy.

However, the biggest difference is the cultural shift and change in user attitudes due to the advent of online banking and popular online marketplaces such as Amazon. Brand association also played a critical role in the success of payments systems, as evidenced in the example of Apple's iTunes platform and PayPal's partnership with eBay \cite{forresthow}.

Over the last two decades, there have been efforts to standardize micropayment protocols by organizations such as the World Wide Web Consortium (W3C), the PayCircle consortium, and the Secure Mobile Payment Services (SEMOPS) project. These projects made valuable contributions, including designs for payment protocols, APIs for payment-enabled applications, and mobile payments solutions, but none of these proposals have thus far been ratified into full standards. These efforts are described in more detail in Appendix.~\ref{sec:appendixstandardization}.

\section{Cryptography-based Systems}
\label{sec:cryptography_based_systems}

Here we present a representative selection of micropayments systems relying on cryptographic mechanisms.

We divide these systems into six categories: in centralized systems, brokers mediate customer-merchant interactions and may even authorize purchases in real-time. In voucher-based systems, customers buy vouchers from brokers which they use to make purchases directly from merchants. Commitment-based systems enable customers to pay using signed commitments, much like paying by cheque. In crypto-token solutions transactions are done using tokens that are computationally infeasible to counterfeit. Some systems rely on probabilistic redemption, i.e. instead of processing multiple small transactions, they probabilistically choose one and inflate the transaction amount accordingly. Peer-to-peer systems adapt several of these solutions for peer-to-peer networks. Bitcoin has emerged as the most popular example of this type. We also discuss proposals to facilitate micropayments using Bitcoin.

\subsection{Centralized Solutions}

\textbf{Chrg-http} \cite{tang1996chrg} invented by Tang and Low in 1996 is not a payment protocol per se, but essentially adapts the Kerberos authentication system to set up a secure channel between customer and merchant. Our customer, Alice, contacts a centralized broker who verifies her identity and issues her credentials to communicate with a merchant. Alice makes purchases and the merchant maintains a running balance and bills her at periodic intervals, thereby amortizing transaction costs. Chrg-http was implemented on the Mosaic web browser.

\textbf{NetBill} \cite{sirbu1995netbill}, developed at Carnegie Mellon University in 1995 in partnership with Visa, is a debit-based system for purchasing digital content.

The process flow is as follows: a NetBill server maintains accounts for customers and merchants. Prior to shopping, Alice charges her account by transferring funds into it. The server issues her credentials consisting of a unique user ID and a public/private key pair. An adaptation of Kerberos is used to authenticate communication between customer and merchant.

To purchase a digital item Alice clicks a button on the website, thereby contacting the merchant with her ID and a request for a price quote. The merchant verifies the ID and responds with a price offer. If Alice approves, she sends an acceptance message, resulting in provisional delivery, i.e. the merchant encrypts the item and sends it to Alice. Delivery of the decryption key, however, is conditional upon payment.

Alice next prepares and digitally signs an electronic payment order (EPO) which she sends to the merchant. The merchant appends the decryption key to the EPO, signs the whole package to endorse it, and forwards it to the NetBill server. The server checks if the details are in order, and if Alice's account has sufficient credit, it authorizes the payment. The payment amount is deducted from Alice's account and credited to the merchant, and the transaction is logged.

The server returns a signed copy of the receipt to the merchant with decryption key attached. The merchant forwards a copy to Alice who can now decrypt her purchased item.

The protocol may appear complicated but discourages fraud and facilitates dispute resolution. The EPO includes timestamps, customer and merchant IDs, and identifiers and hash fingerprints of the digital goods being purchased. Alice only signs the EPO after she has received the goods, thereby ensuring delivery. If the merchant reneges on his commitment after the payment is cleared by withholding the decryption key, Alice may contact the NetBill server directly for it.

NetBill has several advantages. No credit card numbers are sent over the Internet. Alice may maintain multiple NetBill accounts and use pseudonyms to protect her privacy. All transactions are handled within the NetBill system and there are no inter-institution clearing costs. Initial transactions from outside the system to fund customer accounts are typically high volume to amortize transaction costs. Merchant earnings are similarly aggregated before they are transferred from his NetBill account to his bank.

However, there are some shortcomings. Alice may be anonymized to the merchant but NetBill is still privy to all her transactions. A relatively large number of messages have to be exchanged for a successful transaction. There is heavy usage of digital signatures which are compute-intensive operations. Proprietary NetBill software is required.

\subsection{Voucher-based Solutions}

In these systems, customers make purchases using digital vouchers obtained from the broker. These systems are intuitive and easy to understand. An advantage of this approach is that the broker need not be involved in transactions. Examples of such systems include Millicent and Foo.

\textbf{Millicent} \cite{manasse1995millicent} was invented by Mark Manasse at DEC in 1995, to facilitate sale of online content such as news articles and stock quotes, etc. It is a debit-based system which aimed to provide transaction fees in the sub-cent range.

Millicent vouchers are merchant-specific. Brokers purchase vouchers in bulk from merchants and sell them to customers in turn. Some brokers may obtain a license from the merchant to produce the vouchers themselves as per demand. Customers wishing to purchase items first contact the broker to purchase vouchers for the particular merchant. Vouchers are managed by wallet software on the customer's machine. Customers maintain accounts with brokers but not with merchants.

A Millicent voucher (referred to as \emph{scrip}) comprises two parts: the first consists of identifiers, such as scrip ID and merchant ID. The body contains information pertaining to the voucher, such as its value and expiration date. To certify the voucher as genuine, the creator of the voucher concatenates the body of the voucher together with a master secret credential and uses a one-way collision-free hash function to generate an authenticator which is then appended to the voucher.

To initiate a purchase, Alice sends a content request to the merchant along with a voucher for payment. She certifies the request by concatenating it with her voucher and a secret credential she shares with the merchant, and using a hash function to generate an authenticator which she affixes to the request. The merchant checks if her request and voucher are genuine and then compares the voucher against a database of used vouchers to confirm that it has not been used before. He then dispatches her purchased content. Leftover change is issued in the form of a new voucher for the change amount.

Millicent's advantages include the fact that it only uses hash functions which are considerably more lightweight than public key cryptography. The broker does not have be online during transactions as the merchant does verification at his end. Millicent also offers partial anonymity in that the merchant need not know the identify of the customer. However the broker maintains a record of all sold vouchers.

However, there are also disadvantages. For one, vouchers are merchant-specific and the system is best suited for long-term customer-merchant relationships.There are also potential dispute resolution issues as only the creator of the voucher can verify it as genuine, since he alone possesses the master credential used to certify it. Non-repudiation is not possible without public-key cryptography. Furthermore, there needs to be a process enabling customer and merchant to share a secret credential which the merchant later uses to verify the customer's purchase requests.

Millicent was briefly trialled in the United States in 1997. An integration of the Millicent payment system with the Minstrel Push system is described in \cite{puhrerfellner2000implementation}.

A related scheme from Foo and Boyd \cite{foo1998payment} inverts Millicent's protocol for greater efficiency and easy implementation. In their scheme, Alice visits the merchant's website where she can download pre-encrypted copies of the content she is interested in purchasing along with payment vouchers. She then pays for the vouchers at the bank at which point the bank provides her with a decryption key for the goods. The advantage of this scheme is that the burden for processing transactions shifts completely onto the bank and minimal upgrading is required at the merchant's end.

\textbf{Netcents} \cite{poutanen1998netcents} by Poutanen et al. overcomes merchant lock-in by using \emph{floating scrips}, i.e. signed vouchers which may be passed from merchant to merchant, but valid only with one at a time. Vouchers are customer-specific, issued to them by banks, and contain a balance which customers can spend progressively making purchases from different merchants.

To make a payment, Alice sends the merchant a certified electronic payment order and a voucher. To pay a second merchant using the remaining balance on her voucher, Alice requests him to contact the first merchant directly. All three parties then engage in a protocol to transfer ownership of the voucher to the second merchant with an updated balance.

Netcents relies heavily on digital signatures and intensive communication. Each voucher also bears a signed history of its past which is visible to merchants. A proposed solution is to use blinding mechanisms to hide this history.

The influence of voucher-based schemes is clearly evident in peer-to-peer schemes such as PPay and Bitcoin (Sec.~\ref{sec:peer-to-peer_systems}) where the basic currency units are vouchers which are arbitrarily divisible and float from owner to owner.

\subsection{Commitment-based Systems}

In these systems customers make purchases using signed promissory notes from customer to merchant, which merchants encash at the bank at regular intervals. The bank also issues customers and merchants with credentials (ID and public/private key pair) enabling them to engage in transactions.

\textbf{Agora} \cite{gabber1996agora}, a credit-based system invented by Gabber and Silberschatz in 1996, is a representative example.

To initiate a purchase Alice requests a price quotation from the merchant. The merchant responds with the quote, which includes the price of the item, a unique transaction ID, merchant ID certified by the bank, and the merchant's public key. Alice verifies that the merchant ID is current and  the price is correct. She prepares a purchase order, which includes her certified customer ID and public key, the transaction ID, and the agreed item price. She digitally signs this order before sending it, thereby committing to the purchase of the item as specified by the merchant.

The merchant verifies that Alice's ID is authorized by the bank and the signature on the purchase order is valid, and then dispatches the item. At the end of the billing period, the merchant submits the transaction messages to the bank as proof of transaction. The bank debits the corresponding amount from Alice's account and credits it to the merchant.

Banks can facilitate partial anonymity by issuing customers with aliases.
The unique transaction ID protects against replay attacks and double-charging, and the use of digital signatures prevents both parties from altering messages later to claim a different price, as well as protects communication over insecure channels. The authors recommend embedding a hash fingerprint of the purchased item in the exchanged messages for dispute resolution purposes.

In the possibility that a customer (or a thief who has stolen customer credentials) may make large transactions and not pay later, banks can mandate a limit for customer purchases, exceeding which the bank is required to explicitly authorize future transactions. Banks can also alert merchants by periodically broadcasting lists of revoked credentials.

The Agora protocol is remarkably lightweight. The authors develop a Java applet which piggybacks transaction messages onto regular HTTP communications between client and server.

\textbf{MiniPay} \cite{herzberg1997minipay} developed in 1997 by Herzberg and Yochai innovates on the basic Agora protocol by incorporating Internet service providers (ISPs) as brokers and clearing houses. This idea has precedent in the example of premium telephone numbers used to pay for phone services, such as voting for TV shows and adult chat services.

Customers and merchants maintain accounts with their ISPs, which also issue credentials and specify spending limits. Merchants embed pricing information for items in HTML links. Customers make purchases using software wallets which generate signed commitments in a convenient `click-and-pay' manner. At the end of the billing period, the merchant submits purchase proofs to his ISP which consolidates various payments and settles outstanding accounts with other ISPs.

MiniPay was successfully trialled and set to launch as a commercial solution by IBM, but this did not materialize.

\subsection{Crypto-token Solutions}

These systems use cryptography to mint unique non-forgeable tokens to use in transactions. There are two main approaches to generate tokens, both relying on hash functions.

\subsubsection{Hash structures}

These systems derive from Lamport's use of hash chains as one-time passwords \cite{lamport1981password}, also the foundation of Haller's S/KEY protocol \cite{haller1995s}.

A representative system is \textbf{PayWord} \cite{rivest1997payword} developed in 1997 by Rivest and Shamir. PayWord is a credit-based protocol.

In the PayWord ecosystem, the broker issues Alice a certificate validating her as an authentic customer. Alice then mints `paywords', which are payment tokens she will use to pay the merchant. She picks a random seed value which is then repeatedly hashed using a collision-resistant one-way hash function (like SHA1) to generate a chain.

To initialize a purchase, Alice sends a digitally signed commitment to the merchant with information regarding the purchase, such as merchant ID, her customer certificate, and the very last value, or `root', of the hash chain. Computing a signature over this root value essentially certifies the chain and bootstraps the payment process. Each preceding hash value is considered a valid payment token.

The merchant verifies the certificate and signature and authorizes the sale. Alice then sends him successive tokens, an amount corresponding to the price of her item. The merchant hashes the first token he receives to check if it matches the signed root in the commitment message. Each token is likewise checked to verify that it hashes to the previously received token. The one-way nature of the hash function ensures that the merchant can only traverse the chain in the reverse direction, i.e. he can verify tokens, but he cannot generate new ones himself. And due to the one-way nature of the hash function, it is computationally infeasible to forge tokens.

When payment is complete, the merchant sends Alice her item. He sends the commitment message and the final received token to the bank which verifies transaction details and the integrity of the hash chain, and debits Alice's account by the corresponding amount, crediting it to the merchant.

Like Millicent, this scheme is fast and lightweight since hash operations are orders of magnitude faster than digital signature operations and is ideally suited for long-term customer-merchant relationships.


Receiving change is not straightforward and may require a reverse PayWord transaction from merchant to customer. However, Alice may vary denomination of individual paywords in consultation with the merchant and certify the decision by scripting it in the purchase commitment. PayWord can also be used in a debit-based scenario where brokers generate and sell paywords to customers who then use them to shop.

Similar schemes using hash chains were developed independently by multiple parties including Pederson \cite{pedersen1997electronic}, \emph{micro-i}KP \cite{hauser1996micro}, and notably the NetCard project \cite{anderson1997netcard} which attempted to integrate this solution with existing banking infrastructure.

Researchers have also innovated further on PayWord. Kim et al. \cite{kim2003pay} describe a mechanism enabling a customer to transact with multiple merchants with a single hash chain operation. \textbf{PayTree} \cite{jutla2012paytree} amortizes signature costs by integrating PayWord chains with Merkle trees, so that one signature certifies multiple chains at once. The tree structure also opens up other interesting possibilities: different chains can be used to transact with different merchants concurrently, and chains may be efficiently initialized with various token denominations.

\textbf{NetPay} \cite{dai2007netpay} adapts PayWord for decentralized scenarios. In this case banks issue customers with wallets of payword tokens which they use to purchase items. The first merchant verifies tokens received from the customer by directly contacting the bank. The next merchant however contacts the first merchant to verify token he receives, and so on and so forth for other merchants. The bank therefore does not need to be involved every time the customer interacts with a new merchant.

\subsubsection{Hash Collisions}

These schemes exploit the inherent difficulty of finding hash collisions, i.e. two input values which map to the same 160-bit hash output value. As per the `birthday attack' finding a single hash collision requires, on average, hashing through $1.2\times2^{160/2}$ values, which entails not just immense computation effort and time, but also enormous storage requirements (all input/output values have to be stored and searched to identify a collision).

Verifying a collision, however, is extremely easy: one just has to hash the input strings and confirm if the outputs match. This sharp asymmetry in generating and verifying collisions can be exploited by using partial hash collisions as tokens.

This is the approach taken by MicroMint \cite{rivest1997payword}, invented by Rivest and Shamir in 1997, a debit-based system which uses \emph{k}-way hash collisions as payment tokens.

A \emph{k}-way collision is a set of \emph{k} distinct input strings which yield the same partial hash output, i.e. the outputs agree for a specific number of bits. Different combinations of these input strings are packaged into tokens which customers use for transactions. Merchants check the authenticity of tokens when accepting them by verifying the collisions, and later redeem them with the broker. It is impractical to counterfeit these tokens but they can easily be duplicated. To prevent double-spending, brokers only redeem spent tokens once.

We describe here the intuition of how collisions are generated: the broker sets up a number of storage bins to classify input strings based on their hash output, such that strings which create partial hash collisions will end up in the same bin. The authors anticipate that tokens are minted over a month-long period using special-purpose hardware optimized for hash computations, and it is highly likely that many bins will end up with multiple strings. The broker then packages strings in separate bins into individual tokens.

This system calls for substantial initial investment on the part of the broker. Researchers have since proposed optimizations for the minting process \cite{wheeler1996micromint} \cite{wagner1997parallel} \cite{jakobsson1999proofs} and discussed implementation concerns \cite{ramzan2000preliminary} \cite{van2002practical}. Some prototype implementations have also been described \cite{hassler1997mimi} \cite{burstein1998implementation}.

MicroMint has not inspired many derivative schemes, but the notion of using hash function collisions has proved influential in other domains and inspired the notion of proof-of-work schemes \cite{dwork1993pricing} \cite{jakobsson1999proofs} which force a party to undertake computational effort in return for some privilege. Hashcash \cite{back2002hashcash}, a prominent example, was originally proposed to limit spam email and denial-of-service attacks and has since been adopted by Bitcoin and other cryptocurrencies.

\subsection{Probabilistic Audit and Redemption Mechanisms}

These systems reduce processing costs by probabilistically selecting and processing individual micropayments from a set of many such that the odds of identifying fraud or fair compensation are maximized.

\subsubsection{Probabilistic Auditing}

Jarecki and Odlyzko \cite{jarecki1997efficient} attempt to bridge the gap between solutions relying on real-time availability of the broker, like NetBill, and those which necessitate only periodic access at the end of the billing period, examples like PayWord and MicroMint. The goal is to reduce communication overhead while still detecting if the customer tries to cheat or exceed her credit limits.

The solution is for merchants to accept customer payments and use a probabilistic mechanism to determine whether or not to forward the transaction to the bank in real-time. The frequency with which payments are sent to the bank are calculated as a function of the monetary values of the transactions and the amount of risk the bank is willing to take. As transactions grow larger, therefore, the merchant contacts the bank more frequently, whereas, for low value amounts, typical of micropayments, the contact is much less.

Probabilistic auditing can simply be grafted on top of existing systems, and indeed has been proposed as extensions for the NetCents and Agora solutions.

\subsubsection{Probabilistic Redemption}

Wheeler initially floated the idea that \textbf{weighted bets} could efficiently amortize transaction costs \cite{wheeler1997transactions}. A simple illustration: for Alice to make a payment of 37 cents currently requires a transfer of at least four coins. Furthermore, lets assume that Alice only possesses dollar bills. One solution is to toss a 100-sided die; if it yields a number between 1 and 37, Alice pays a dollar to the merchant, and if not, she gives him nothing. If this transaction is repeated often enough, the average payment Alice makes to the merchant is 37/100 dollars, i.e. 37 cents.

\textbf{Rivest's lottery-based scheme} \cite{rivest1997lottery}, developed in 1997, extends this notion to PayWord. In this case, the bank issues Alice with a book of ``lottery tickets" essentially consisting of a PayWord chain with a specified lifetime. As with a PayWord purchase, Alice uses successive values in the chain as payment tokens and the merchant verifies each token accordingly before sending over the corresponding items.

After the lifetime of the chain expires, the bank announces that one individual lottery ticket in each book is a ``winning'' ticket. If Alice has transferred this ticket to the merchant in the course of her purchase, she is obligated to pay. In this case, the merchant presents the winning ticket to the bank which debits Alice's account for the full value of her purchase and credits the amount to the merchant's account. However, if Alice retains the ticket (i.e. she hasn't spent it), she pays nothing. In the long run, when thousands of such transactions have happened, the probability is very high that the amount Alice is charged will converge to the actual amount she owes the merchant. Processing costs are minimized because the bank only has to process winning tickets.

This scheme suffers drawbacks: the bank has to organize the lotteries, circulate details of winning tickets, and the merchant can only be paid after the lottery concludes. Rivest extends the protocol to address these concerns and proposes a fair mechanism allowing Alice and the merchant to decide among themselves whether a ticket is a winning ticket or not \cite{rivest1997electronic}.

Ostrovsky and Lipton propose a similar scheme to minimize  the bank's involvement \cite{lipton1998micro}. Here Alice and the merchant both exchange roots of pre-computed hash chains prior to the transaction, as well as cryptographic commitments to the seed values from which the chains are generated. In each round, both parties produce their tokens which are then XOR-ed to yield the output for the round. Since both tokens are pseudo-random values, the process is analogous to tossing a coin. Alice pays the merchant depending on the result of the toss.

Rivest revisited the topic with Micali in 2001 \cite{micali2002micropayments} \cite{rivest2004peppercoin} to present \textbf{Peppercoin}, a non-interactive version of the lottery scheme. In this case, winning tickets are selected by a cryptographic process: the merchant digitally signs a token and checks if the signature result is less than a pre-determined value. It is important that a deterministic digital signature scheme be used (such as RSA), so that the merchant can convince the bank that a ticket is a winning ticket. Given the cryptographic nature of digital signatures, neither Alice nor the merchant can game the system in their favour. Furthermore, to protect Alice from being charged too much at once, Peppercoin brings in banks as intermediaries, or buffers, who pay the merchant the inflated amount but charge her only the aggregated value of the payments she has made.

Peppercoin launched as a commercial system in 2001, starting services in 2003, but closed in 2007.

\subsection{Peer-to-peer Systems}
\label{sec:peer-to-peer_systems}

The emergence of large-scale peer-to-peer (P2P) networks in the early 2000s necessitated research on payment solutions for this new paradigm. There were two main motivations: first P2P networks suffered from the free-rider problem, i.e. certain peers would solely use network resources without contributing any themselves. One solution is to deploy a metering or micropayment solution to incentivize participation and ensures fairness in the network \cite{golle2001incentives} \cite{ranganathan2003share}.

The second reason is commercial. Networks like Napster and Kazaa suffered immense backlash and heavy litigation from the recording industry for freely sharing pirated content, leading researchers to examine possibilities for transitioning these networks to legal marketplaces.

\textbf{PPay} \cite{yang2003ppay}, presented by Yang and Garcia-Molina in 2003, is one of the earliest and most influential P2P micropayment systems. PPay is a debit-based protocol that adapts several innovations from past systems to a P2P environment.

In the PPay ecosystem, a broker issues coins to Alice which are essentially certified vouchers of fixed denomination, bearing identifying information such as a unique serial number and Alice's identity.  To make a payment, Alice prepares a reassignment message, consisting of identity of the merchant, the coin's information, and an assigned sequence number which increments every time the coin changes hands. Alice signs this message and sends it to the merchant who then becomes the official `holder' of the coin (as opposed to Alice who is the `owner' of the coin).

The merchant may then use the coin, in the role of a customer, to make a purchase from another merchant. In this event, he sends Alice a reassignment request, i.e. a digitally signed message with details of his last transaction with Alice and the identity of the new merchant. Alice prepares and sends back to both a reassignment message with the identity of the new merchant and incrementing the assigned sequence number. Alice also logs the reassignment request, as evidence that the first merchant relinquished the coin. The broker is only approached when users want to cash out of the system.

PPay has considerable strengths: first, broker involvement is considerably reduced as coin owners manage security of the coin. All messages are digitally signed by all parties, thereby enabling forensic analysis later on to identify any instances of double-spending or fraud.

However, there are also some weaknesses: coin owners are required to be online to facilitate transactions. A solution, the authors suggest, is to enable coin holders to issue reassignment messages themselves which are `appended' to the coin, giving rise to the notion of ``layered'' coins. Each layer is a new reassignment message, which can be ``peeled'' back later on to validate the provenance of the coin by verifying signatures and sequence numbers. Another alternative is to allow brokers and coin-owners to conduct probabilistic audits of transactions and reassignment requests.

PPay has proved immensely popular and several schemes have built on its basic features. A notable example is \textbf{WhoPay} \cite{wei2006whopay}, by Wei et al. which provides revocable anonymity to coin holders (not owners) by employing group signatures. The identity of the current coin holder is concealed as one within a group but, in exceptional circumstances, may be unmasked by cooperation of the broker and a trusted authority. Coins are not represented by serial numbers but by public keys, such that the owner of the corresponding private key is the holder of the coin. For each transaction, the merchant generates a new public/private key pair, and the original owner signs the coin along with the new public key, to designate the reassignment. The authors also suggest a public log for all transactions which peers can check in real-time to detect double-spending.

\textbf{FairPeers} \cite{catalano2005p2p} \cite{ruffo2007fairpeers}, by Catalano and Ruffo, adapts PPay for selling copyright content. A Copyright Granter entity, issues digital certificates testifying to the authorship of a file. A merchant offers to sell the file. To purchase it, Alice will need two different coins, one sent to the merchant and the other to the original author of the file. Unfortunately, in this scenario the requirement that authors always be online for payments can be problematic. Some solutions are considered in \cite{ruffo2005scalability}.

Another influential system, \textbf{Karma} \cite{vishnumurthy2003karma}, addresses the free-rider problem in a decentralized manner. Coins are replaced with the concept of \emph{karma}. A new node, say Alice, entering a P2P network is associated with a set of peers (called the \emph{bank-set}) which act as a semi-trusted authority, tracking the resources Alice consumes and contributes, and maintaining a record replicated across all the peers. When Alice makes a transaction with a merchant by transferring him an amount of karma, all the peers in her bank-set send messages to all the peers in the merchant's bank-set, testifying to Alice's balance of karma, thereby validating the transaction. The merchant's bank-set uses a majority voting protocol to confirm the transaction is in order, and the merchant then sends Alice her purchased items. This system assumes that the majority of peers in the network are honest.

A drawback of Karma is that bank-set peers need to be online to validate transactions. This limitation is addressed by \textbf{Offline-Karma} \cite{garcia2005off}, where each reassignment adds a new ``layer'' of provenance to a coin. Coins have a fixed lifetime after which the layers are peeled back to check for double-spending and fraud. This function is undertaken by a distributed set of nodes, known as the reminters, who then affix a multisignature to the coin, re-certifying it for use.

\textbf{CPay} \cite{jia2005new} takes a different route to resolving the peer-availability problem. Instead of relying on multiple peers to authorize a transaction, the customer runs a function to select one peer from a set of online peers who have been designated Broker Assistants by the broker. The Broker Assistant then verifies the transaction.

Other P2P schemes in this vein include Zuo and Li's Fair Exchange File Market \cite{zuo2005constructing}, which describes how to integrate a payment solution between customer and merchant in a BitTorrent-type scenario where file pieces are distributed among multiple peers and have to be retrieved. Attacks on this system are described in \cite{piva2007strand} \cite{piva2009regarding}. P2P-NetPay \cite{dai2005off} adapts the PayWord-based NetPay protocol described earlier for P2P networks. A prototype implementation is described in \cite{chaudhary2009p2p}.

The most popular and influential system in this category though is undoubtedly Bitcoin. Developed by Satoshi Nakamoto in 2008, \textbf{Bitcoin} has achieved mainstream success and is currently the world's leading cryptocurrency with a market cap of around \$11 billion \cite{coinmarketcap}. Bitcoin brings together and harmonizes several of the innovations we have discussed thus far: the currency units, bitcoins, are essentially floating vouchers. Users maintain a public/private key pair. A hash of the public key is considered the user's `Bitcoin address'. To make a payment the user digitally signs a transaction message using her private key and the balance of bitcoins is transferred to the receiver's Bitcoin address.

Bitcoin's most important innovation is the blockchain, a distributed public ledger which records all transactions on the network and enables peers to detect double-spending. This is literally a chain of blocks, each of which contains recent transactions. A decentralized set of miners is responsible for the creation of new blocks and they compete among themselves to solve a computationally difficult puzzle. The winner creates and appends the next block in the chain. This leads to Bitcoin's second innovation, a monetary reward that incentivizes miners to maintain the Blockchain.

A detailed description of the Bitcoin payment system goes beyond the scope of our paper (interested readers are directed to \cite{tschorsch2015bitcoin}). We note in brief some key properties of the system. Most importantly, the network is distributed and trustless, i.e. no centralized broker or bank is required to authorize transactions or check for fraud. Bitcoin users transact using pseudonyms which confers a degree of anonymity, but research has evolved methods to attack it \cite{meiklejohn2013fistful}. Bitcoin transactions generally include a transaction fee (typically in the order of cents or lower) to incentivize miners to include the transaction in the blockchain. An important distinction here is that transaction fees are not a function of payment amount but rather depend on the amount of data in the transaction.

Bitcoin is described as a general payments system but its relatively low transaction fees qualifies it as a micropayments system in its own right. Indeed several commercial micropayments systems (some of which are described in Sec.~\ref{sec:commercial_solutions}) already use the Bitcoin network to process payments.

However, researchers have innovated mechanisms to further drive down processing costs using Bitcoin. We consider some of these next.

\subsection{Enabling Micropayments on Bitcoin}

The most straight-forward approach to limit transaction fees is to amortize transaction costs by conducting \textbf{off-chain transactions} and only settling their accounts on the Bitcoin network. The easiest way to achieve this is by introducing a broker like ChangeTip or Coinbase to maintain accounts for customer and merchant on their own system. However, this strategy is not recommended: Bitcoin exchanges have a notoriously poor track record of protecting customers' funds  \cite{moore2013beware}. We discuss this point in more detail in Sec.~\ref{sec:security_challenges}.

Protocols have been proposed to enable off-chain transactions without a trusted third party \cite{bitcoinpaymentnetworks}. Referred to as \textbf{micropayment channels}, these solutions rely on the blockchain to resolve payment disputes. We consider an example: two users wishing to transact set up a channel by depositing funds into a special transaction which is then stored on the blockchain. Access to the funds is prohibited without authorization of both parties. The users then make multiple low-volume transactions to each other privately off-network which effectively redistribute this deposit among themselves. To terminate the channel and redeem their funds, either party can broadcast the last transaction they exchange on the Bitcoin network. As a safety mechanism, the initial transaction bears an expiry time, after which, if no payments have been made, the deposit is automatically refunded.

We briefly describe some basic micropayment channels:

A \textbf{uni-directional channel} permits micropayments to be made strictly in one direction only, that is from one user to another.
Only the sender has to make the initial deposit onto the blockchain. Each micropayment then increases the receiver's share of the deposit. The amount of the deposit sets the upper limit on the funds the sender can transfer to the receiver for the session.

Interestingly, timers can extend support for \textbf{bi-directional micropayments}. The timer determines the minimum block height at which the transaction is accepted into the blockchain and setting the appropriate timer value ensures that one transaction takes precedence over others. In this case, when the sender makes a payment and the receiver wishes to make one back, the sender can broadcast another transaction reducing the balance of his last transaction by the appropriate amount. Decrementing the timer value ensures the second transaction with the amended balance is included in the blockchain and the first is not.

Poon and Dryja propose \textbf{Lightning Channels} \cite{lightningnetwork} that supports bi-directional payments with infinite lifetime. The channel has an active transaction representing the current balance of both users and a list of revoked transactions. Sending a micropayment entails revoking the active transaction and replacing it with a new one that represents the new balance of both parties. If a user circulates a previously revoked transaction on to the Bitcoin network, the other user has a time-period to broadcast a penalty transaction and acquire all the bitcoins in the transaction. Either party can close the channel by broadcasting the latest active transaction or both parties can co-operate to settle the final balance.

Other approaches are being developed independent of off-network micropayments channels: for instance, \textbf{MicroPay} \cite{pass2015micropayments} provides a version of probabilistic payments system that is compatible with Bitcoin. Recent research proposals have also attempted to extend ZeroCash, a new and more privacy-conscious cryptocurrency, to support micropayment channels and probabilistic payments in the offline setting \cite{zcmicropayments}.

\begin{table}

\protect\caption{Classification of Cryptographic Systems \label{tab:Crypto-System-Properties}}

\begin{centering}
\begin{tabular}{cccccrrl}
\hline
Scheme  & Strategy  & \multicolumn{3}{c}{Properties} & \multicolumn{3}{l}{Payment}\tabularnewline
\hline
\hline
 &  & \rotatebox{90}{Anonymity}  & \rotatebox{90}{Guaranteed Delivery }  & \rotatebox{90}{Non-repudiation }  &  & \rotatebox{90}{Pre-paid} & \rotatebox{90}{Post-paid}\tabularnewline
\hline
Chrg-http  & Centralized  &  &  &  &  & $\CIRCLE$ & \tabularnewline
\hline
NetBill  & Centralized  &  & $\CIRCLE$ & $\CIRCLE$ &  & $\CIRCLE$ & \tabularnewline
\hline
Millicent & Voucher-based  &  &  &  &  & $\CIRCLE$ & \tabularnewline
\hline
Netcents  & Voucher-based  &  &  &  &  & $\CIRCLE$ & \tabularnewline
\hline
Agora  & Commitment-based  & $\LEFTcircle$ & $\LEFTcircle$ & $\CIRCLE$ &  &  & $\CIRCLE$\tabularnewline
\hline
Mini-pay & Commitment-based  & $\LEFTcircle$ & $\LEFTcircle$ & $\CIRCLE$ &  &  & $\CIRCLE$\tabularnewline
\hline
Payword  & Crypto-tokens  &  &  & $\CIRCLE$ &  &  & $\CIRCLE$\tabularnewline
\hline
PayTree  & Crypto-tokens  &  &  & $\CIRCLE$ &  &  & $\CIRCLE$\tabularnewline
\hline
NetPay  & Crypto-tokens  &  &  & $\CIRCLE$ &  & $\CIRCLE$ & \tabularnewline
\hline
 MicroMint & Crypto-tokens  & $\CIRCLE$ &  &  &  & $\CIRCLE$ & \tabularnewline
\hline
Peppercoin & Probabilistic &  &  & $\CIRCLE$ &  &  & $\CIRCLE$\tabularnewline
\hline
\end{tabular}
\par\end{centering}

\end{table}

\subsection{Discussion}

There is a clear evolutionary trend in the systems we have thus far examined. Early systems like Millicent, Agora, PayWord, and MicroMint, showcase a variety of broad innovative approaches towards micropayments. There is considerable cryptographic innovation with an emphasis on security and efficiency. Later systems, such as P2P systems are more application-oriented and tend to synthesize these different approaches.

We also observe a visible shift in design priorities and limitations as technology advances and infrastructure improves. Early schemes like Millicent and PayWord went to great lengths to minimize usage of digital signatures and transaction latency whereas P2P schemes have no such restrictions.

\section{Commercial Solutions}
\label{sec:commercial_solutions}

We examine a selection of commercial micropayment systems that are currently in use, or have had significant impact.

We divide these systems into four categories: in pre-paid systems customers deposit funds into their accounts which they then progressively spend. Several systems facilitate payments in various ways, either by allowing an existing payment method to be used in a new area, or enabling customers and merchants to manage their payments and purchases better. Other systems amortize transaction costs by aggregating multiple payments into smaller numbers of transactions. Many of these systems support both small and large payments, and the target markets and benefits they provide differ considerably.

\subsection{Pre-paid Systems}
\label{sec:Pre-paid-Systems}
These systems involve a user making an advance payment to the payment provider via cash or credit/debit card. This payment is converted into funds inside the system which can be used to pay participating merchants.

\textbf{Paysafecard} \cite{paysafecard}, launched in 2000, is a system based
around pre-paid scratchcards. Customers buy scratchcards in advance
that have a value ($\euro5,\euro10,\euro25,\euro50$ or $\euro100$)
and a 16 digit PIN. When the customer makes a payment she
enters the PIN to authorise the payment. If there
is not enough balance on the card for the whole payment she can enter
additional PINs and use up to 10 cards.

Merchants receive monthly payments from paysafecard that combine all
of the transactions made for that month, thereby reducing processing costs. Transaction fees depend on
the location and business area of the merchant.

\textbf{PayPal} \cite{paypal}, established in 1998, is an account-based system
where users deposit and withdraw money via credit/debit card.
Transactions are made in real-world currencies and merchants
pay fees for every transaction that occurs.

PayPal has special merchant accounts for micropayments. These accounts
function like normal merchant accounts, but have a different transaction
fee structure, charging \$0.05 plus 5\% of the transaction . Normal
merchant accounts charge \$0.30 plus 2.9\% of the transaction, with
the possibility for merchants with large transaction volumes or non-profit
status to negotiate a fee as low as \$0.30 plus 2.2\% of the transaction.

PayPal also offers direct carrier billing for customers whose mobile
service providers are part of the PayPal carrier network. The PayPal carrier network has a wide coverage worldwide due
to PayPal's acquisition of direct carrier billing company Zong and
partnership with Deutsche Telekom.

\textbf{Flattr} \cite{Flattr}, launched in 2010, is a system enabling
users to support content creators such as artists, musicians or writers. The user
sets a monthly budget which is pre-paid into the system by bank transfer or credit/debit card each month. The user also maintains a list of people they choose to support. Each month the monthly budget is divided
equally among the people on the list.

As only one payment is taken each month, only one transaction cost
is accrued, regardless of the number of people on the list. Similarly,
content creators are given one monthly payment combining all of
the payments sent to them by users during that month. This reduces
the fixed portion of the transaction fees to one fee per user and
one fee per content creator, a significant reduction when users are
paying small amounts to many content creators.

Flattr also reduces mental transaction costs, as users do not
have decide how much to pay each creator, instead simply
adding people to the list when they see something they like, and removing
people they no longer wish to support.

\textbf{ChangeTip} \cite{ChangeTip}, founded in 2013, is designed
to allow users to make small one-off payments to people, businesses and organizations they wish
to support. Users deposit money into their accounts
via Bitcoin or credit/debit card. Transfers of any amount from one user to another
are free and withdrawals are charged a small transaction fee, set
at different levels for dollars and bitcoins.

As with Flattr, ChangeTip amortizes processing
costs as one large transaction deposits money into the system
and enables multiple small payments to be made without further
costs.

\textbf{Click and Buy} \cite{clickandbuy}, founded in 1999, is an account-based system where customers deposit money via credit card and bank
transfer. A 3.9\% transaction fee is levied for credit card deposits,
with bank transfer deposits being free of charge. Merchants are charged
both a fixed fee per transaction and a percentage of their total revenue. The amount charged is based on the average transaction
amount, with merchants who generally receive smaller amounts per transaction
having a smaller fixed fee and a larger percentage than merchants
who generally receive larger amounts per transactions.

Merchants can opt to be paid by Click and Buy on a schedule ranging
from once a day to once every 30 days, with merchants who opt for
longer payment schedules being rewarded by smaller transaction fees.

\textbf{M-Pesa} \cite{mpesa}, founded in 2007 is a mobile phone-based account-based
system where users can deposit and withdraw money through businesses
acting as agents. Users can transfer money both to other users
and non-users.

M-Pesa's largest market is Kenya, where direct transfers to and from
bank accounts are possible. The system is also available
in a number of other countries. Transaction fees are based on transaction
size, with a fixed charge being levied for transfers within particular size categories. Different transaction fees are charged for
transfers to users and non-users.

\textbf{League of Legends} \cite{leagueoflegends}, released in 2009, is one
of the most popular online games and uses a free-to-play model. Revenue
is generated by the sale of in-game upgrades to players. These upgrades
are bought with Riot Points, a currency used only for this purpose.
Riot Points can be purchased with credit/debit cards, PayPal, paysafecard,
bank transfer or using pre-paid cards from a variety of retailers.
Transaction fees are kept low by selling Riot Points in relatively
large blocks, so that the lowest purchase price is between \$2.00
and \$10.00, depending on the purchase method.

As with with most in-game currencies, Riot Points are not convertible
into other currencies, and players are generally prevented from selling
or transferring their points to others.

\textbf{Blendle} \cite{blendle}, founded in 2013, is a news aggregation site
that sells articles using a pay-per-article model. Customers deposit funds
into their Blendle account using a credit card, and then buy access
to articles or entire editions of periodicals. Content
providers are allowed to set a price (currently between \euro0.99
and \euro1.99) for each article they provide, with Blendle charging
a fee of 30\% of this price to the merchant.

Blendle allows customers a refund on any article they have
purchased, with the requirements that the customer requests the refund
within 24 hours and provides a reason for it. This
step is designed to prevent customers paying for articles that
were misleadingly advertised or badly written.

The \textbf{Starbucks card} \cite{starbucks} is an account-based system that
allows customers to buy beverages and food from Starbucks. Customers begin
by buying a physical card with a pre-loaded value. This card has a
unique card number and security code that are stored in a central
database along with the balance on the card. The balance
can be spent at Starbucks stores like a traditional gift card. The card can even be registered online as belonging to a particular user.

A registered card can have additional value added to it in-store or
online, and registered users can also download iPhone and Android
applications that allow the mobile device to be used in place of the
card. Users are also able to transfer value to other registered cards,
allowing the Starbucks card to be used as a form of currency among
users, and use of a registered card is linked to a customer rewards
program.

\textbf{Bitwall} \cite{Bitwall}, is a system designed to allow the purchase
of access to web content. The system is currently in beta stage and
the pricing model is still being determined. At present users can
buy access to one article for \$0.01, 24 hour access to an entire
site for \$0.03 or 3 hour access to the site by advertising the site
on Twitter. Payments are made via Bitcoin, using the current Bitcoin to US dollar exchange rate.

\textbf{Mondex, VisaCash, Proton and Octopus} are systems based around a smart-card that is pre-loaded
with funds and used to make in-person payments. Of the four,
only Octopus \cite{octopus}, launched in 1997, has achieved commercial
success, being used heavily for public transport, car parking charges,
fast-food and vending machine purchases in Hong Kong. Octopus cards
can be reloaded with funds using cash at a range of participating
retailers, and can also be linked with a credit card using the ``Automatic
Add Value Service,'' causing funds to be added whenever the
card reaches zero balance.

Merchants are charged a percentage of each transaction for the use
of Octopus (this percentage can vary between merchants and the factors
that decide the percentage charge are not publicly available) but
no fixed transaction cost.

\subsection{Facilitating Merchants}

These systems improve the ease with which merchants can accept
payments. They generally work as gateways or aggregators, enabling
merchants to collect payments from multiple systems without having
to hold accounts with each, thereby reducing both
expense and effort for the merchant.

\textbf{Mollie} \cite{Mollie}, founded in 2004, is a payment gateway allowing
payments to be made by various means into one account. It does not
provide a micropayment system of its own, but instead allows merchants
to accept payments by credit card, PayPal, Bitcoin, paysafecard and
various bank transfer systems. Fees depend on the payment method used,
with most methods including a fixed fee of \euro0.25.

Previously, Mollie allowed customers to make payments via SMS or by calling
telephone numbers, with the payment amount being added to their telephone
bill. This payment method is no longer offered.

\subsection{Facilitating Customers}

These systems use technology to improve the payment experience for customers.

\textbf{Android Pay} \cite{androidpay}, established in 2015 is an Android
device-based system that can store credit card, debit card and other
card details and enables secure payment via NFC or over the Internet. \textbf{Apple
Pay} \cite{applepay}, established in the US in 2014, is a similar offering based around Apple technology.

Merchants require a payment processor for the underlying card and
are only charged for the card used; Android Pay and Apple Pay do not
add any additional fees. While these systems change the user experience
and the security properties of the transaction, they does not change
the fee structure of the underlying payment method.

\textbf{Paym} \cite{paym} is a system allowing payments between
individuals who are identified by their mobile telephone numbers.
Both parties in a transaction must have bank accounts with participating
UK banks, and both parties must have registered their mobile phones
with the Paym system. The payment sender can then use the Paym mobile
app to make a payment to the recipient, by entering the amount
and the recipient's mobile telephone number.

At present Paym payments bear no charge, but many banks
restrict use of the system with business accounts or levy charges
on transfers to and from these accounts.

\subsection{Aggregation}

These systems combine multiple payments together where possible
to minimize the transaction fees charged by payment processors.

\textbf{iTunes} \cite{itunes}, founded in 1998, accepts standard credit and
debit card payments. Payments are often delayed by a short time (such
as one or two days) to increase the likelihood that the customer will
make further purchases. All of the payments are then lumped together before clearing them to reduce transaction fees.

\textbf{eMusic} \cite{eMusic}, initially founded in 1995 and relaunched in
2004, uses a business model that can be thought of as a combination
of aggregation and pre-paid systems. Users choose a monthly payment
level, which then entitles them to download a number of songs during
that month. One transaction is made by debit/credit card for the monthly
payment, resulting in a guarantee that only one transaction fee will
be paid to the payment processor each month.

This system forces users to commit to buying a certain number of tracks
each month, allowing more aggregation of purchases than may occur
with a more flexible system.

\subsection{Summary}

\begin{table}

\protect\caption{System Properties \label{tab:System-Properties}}

\begin{centering}
\begin{tabular}{ccccc}
System & Target Market & \rotatebox{90}{Pre-paid} & \rotatebox{90}{Anonymity} & \rotatebox{90}{Accepts Cash} \tabularnewline
\hline
\hline
paysafecard & Online markets & $\CIRCLE$ & $\LEFTcircle$ &  $\CIRCLE$\tabularnewline
\hline
PayPal & Online markets &  $\CIRCLE$ & & \tabularnewline
\hline
Flattr & Flattr users &  $\CIRCLE$ &  & \tabularnewline
\hline
ChangeTip & ChangeTip users &  $\CIRCLE$ &  & \tabularnewline
\hline
Click and Buy & Online markets &  $\CIRCLE$ &  & \tabularnewline
\hline
M-Pesa & All payments &  $\CIRCLE$ &  &  $\CIRCLE$ \tabularnewline
\hline
 Points & LoL upgrades &  $\CIRCLE$ &  &  $\CIRCLE$ \tabularnewline
\hline
Blendle & Online articles &  $\CIRCLE$ &  & \tabularnewline
\hline
Starbucks Card & Starbucks products &  $\CIRCLE$ &  &  $\CIRCLE$ \tabularnewline
\hline
Bitwall & Online articles &  $\CIRCLE$ &  & \tabularnewline
\hline
Octopus & In-person payments &  $\CIRCLE$ &  &  $\CIRCLE$\tabularnewline
\hline
Mollie & Online markets &  &  & \tabularnewline
\hline
Apple Pay & Online markets &  &  & \tabularnewline
\hline
Android Pay & Online markets &  &  & \tabularnewline
\hline
Paym & Money transfers &  &  & \tabularnewline
\hline
iTunes & iTunes store &  &  & \tabularnewline
\hline
eMusic & eMusic store &  $\CIRCLE$ &  & \tabularnewline
\hline
\end{tabular}
\par\end{centering}

\end{table}

The commercial systems we have listed all appear to have grown around
a specific problem or application, rather than being formulated to
produce a general micropayment system. The majority of the systems
do not provide anonymity, with paysafecard as the notable exception
(with the possibility for anonymity by a technically skilled user
in the cases of Bitwall and League of Legends Riot Points). Privacy
is generally handled by policy rather than technical safeguards. The
system properties are summarised in table \ref{tab:System-Properties}.

It is also notable that none of these systems rely on novel cryptographic
protocols for properties like non-repudiation of transactions or to
prevent double spending. Instead, all of the systems are account-based,
relying on a trusted third party to authenticate transactions. Even
paysafecard, which doesn't rely on a traditional notion of an account,
requires a central authority to distinguish between valid and invalid
PINs.

\section{Outstanding Challenges and Future Directions}
\label{sec:outstanding_challenges_and_future_directions}

Here we outline key challenges facing design and deployment of micropayments systems. Payments systems today are particularly vulnerable to security threats and we present relevant insights from the Bitcoin experience. We then consider ethical and legal issues that need to be resolved to integrate micropayments successfully into the financial infrastructure.

This is followed by a discussion of cognitive costs and mental models for micropayments. There is very little work done in this domain, and we anticipate that research may enable improved usability for micropayments systems and successful business models. In conclusion we discuss potential deployment strategies for upcoming micropayments systems.

\subsection{Security Challenges}
\label{sec:security_challenges}

Security is a pronounced concern for online payments systems in general. A 2013 report by CyberSource calculates that for online shopping with card in the UK, i.e. the card-not-present paradigm (CPN), the fraud rate dominates and is about ten times higher than for physical credit card fraud \cite{cybersource2013online}. A concurrent study by FICO discovered that the US credit card fraud rate is spiking, surging 17\% over two years \cite{pr2013fico}.

The trend is even more ominous in the Bitcoin community \cite{ali2015bitcoin}. A key reason why Bitcoin's transaction fees are so low is that the systems has no fraud protection or dispute resolution mechanisms. Bitcoin exchanges, marketplaces, and wallet services are routinely hacked, resulting in thefts and losses of hundreds of thousands of customers' bitcoins. Mt. Gox is a most prominent example: the world's largest Bitcoin exchange lost over half a billion dollars worth of bitcoins in an incident, impacting user confidence in the currency itself. One study documents that of 40 Bitcoin exchanges established recently, 18 shut down soon after \cite{moore2013beware}.

In parallel, researchers from Dell indicate a near ten-fold increase in malware designed to steal bitcoins from users' computers, the rate of creation loosely tracking the increase in Bitcoin's own exchange rate \cite{dellmalware}. Some of these even employ keyloggers to crack password-protected wallets. 50\% of these malware successfully bypass most antiviruses.

Micropayments systems are particularly vulnerable because system security is not an isolated feature. Instituting fraud protection mechanisms in a system will almost certainly add to payment processing costs and cut into the broker's profit margins. This increase can be justified for macropayments; for instance, debit card transactions in the US average \$39 per transaction and the interchange fee is about 24 cents per transaction of which 1 cent goes to fraud protection \cite{board2014interchange}). But for very low-value payments, this increase is significant.

A related concern is that adding a fraud protection feature to a system will likely impact system usability and add further cognitive burdens on the user. For example, a user may be amenable to the extra ``hassle" of two-factor authentication for making a macropayment, but for very low-value transactions, the amount might not justify the effort.

Any solution addressing security for micropayments systems will have to harmonize these concerns.

\subsection{Legal and Ethical Concerns}

Micropayment systems, like most real-world applications, will require legislative protection as well. Some scenarios simply cannot be prevented with technology alone (for instance the supply of defective goods in a conventional e-commerce scenario) and others may be independent of technology (such as assigning liability if a payment system suffers losses due to faulty implementation or mismanagement). Legislators need to decide how best to balance interests of users and merchants, and merchants and brokers have to decide if there is a business case to absorb more risk themselves in the interest of gaining market share and consumer confidence.

New payment systems may also introduce the risk of losses beyond what the system can absorb. Bitcoin exchanges have shown that it is possible for a company to suffer losses far in excess of its assets when things go wrong \cite{Moore2013}. This has troubling implications for consumer confidence and is an important concern for both legislators and businesses.

Legislators also have a role in regulating payments systems and preventing customer exploitation. Recent EU legislation has capped interchange fees for using credit and debit cards at 0.3 and 0.2 \% respectively \cite{Interchange2016}. It has also required that card processors provide information to consumers about costs associated with each transaction and a breakdown of the fees. Micropayments systems will require similar oversight.

There are privacy issues as well. Already, concerns have been voiced about companies amassing and monetizing user data \cite{Riederer2011}.
A good example is Octopus (discussed in Sec.~\ref{sec:commercial_solutions}): in 2010, the Hong Kong Privacy Commissioner for Personal Data found Octopus Rewards Ltd. to have breached data protection principles with regard to the sale of customer data to business partners for direct marketing purposes \cite{privacycommisioner}.

As we observed in Sec.~\ref{sec:cryptography_based_systems}-\ref{sec:commercial_solutions}, very few micropayments systems offer the user anonymity from both merchant and broker.  If micropayments are used to purchase individual articles, videos or audio recordings then users' purchase histories can leak sensitive information such as their political inclinations, religious values and sexual preferences.

However, strict anonymity also poses a problem for governments. While many governments might wish to protect the privacy of their citizens, at the same time there will generally be legislation against large anonymous payments, as part of money laundering and anti-terrorism legislation.

Anonymity becomes even more of a concern in the digital-only, `cashless' vision of society emerging in countries like Denmark, Sweden, and Finland \cite{reuters2015denmark}. As digital payments become commonplace, banks and payment processors can blacklist parties for political reasons. A real-world example of a payment blockade occurred in 2010 when US companies refused to process payments sent to Wikileaks \cite{wikileaks2012}.

Discovering ways to allow massively scalable micropayments systems which protect customer privacy and free speech while at the same time preventing tax evasion and terrorism financing is both a legislative and technological challenge.

\subsection{Micropayments and Psychology}

As we noted in Sec.~\ref{sec:background}, payments systems impose mental transaction costs, i.e. the cognitive effort involved in deciding whether an item is worth buying or not, regardless of price. These costs arise due to various reasons: for instance, a large variety of choices can pose a mental bottleneck. As an example, it is simply less mental effort for a user to buy a whole newspaper for a set amount than to compare the anticipated merits and prices of individual articles. This may explain the appeal of flat fees.

In an influential position paper on the topic, Szabo \cite{szabo1999micropayments} contends that as prices go down, these mental costs tend to dominate over technological costs, they set the effective lower bound on pricing of goods, and therefore, may play a determining role in the adoption of micropayments systems.

Szabo makes some recommendations for systems developers. First, rather than focus on technological innovation alone, it is imperative to recognize mental transaction costs. Technology may then be applied to alleviate these costs.

One proposed approach is to employ metaphors to simplify the mental effort involved in making choices. We present here an example employed by ChangeTip (described in Sec.~\ref{sec:commercial_solutions}) which presents users with payment options labelled as real-world items of similar cost (shown in table \ref{tab:ChangeTip-Amounts}). For example, a user can choose to reward a party by ``buying them a coffee'', press the requisite button, and the system will transfer the corresponding amount of money. Users may also define their own custom amounts on ChangeTip.

\begin{table}
\protect\caption{ChangeTip Amounts\label{tab:ChangeTip-Amounts}}

\begin{centering}
\begin{tabular}{|c|c|c|}
\hline
Label & Amount & Currency\tabularnewline
\hline
\hline
Beer & 3.50 & USD\tabularnewline
\hline
Buck & 1.00 & USD\tabularnewline
\hline
Cent & 0.01 & USD\tabularnewline
\hline
Cerveza & 3.50 & USD\tabularnewline
\hline
Coffee & 1.50 & USD\tabularnewline
\hline
Cookie & 1.50 & USD\tabularnewline
\hline
Dime & 010 & USD\tabularnewline
\hline
Dollar & 1.00 & USD\tabularnewline
\hline
Donut & 0.35 & USD\tabularnewline
\hline
Euro & 1.00 & EUR\tabularnewline
\hline
Gold-star & 0.50 & USD\tabularnewline
\hline
High-five & 5.00 & USD\tabularnewline
\hline
Nickel & 0.05 & USD\tabularnewline
\hline
Pie & 3.14 & USD\tabularnewline
\hline
Pint & 3.50 & USD\tabularnewline
\hline
Quarter & 0.25 & USD\tabularnewline
\hline
Quid & 1.00 & GBP\tabularnewline
\hline
\end{tabular}
\par\end{centering}

\end{table}

Branding and quality control is another approach. If a brand consistently delivers good quality for money, users may be more inclined to trust its product and exert less mental effort in choosing to buy it. One way to implement this would be the strategy taken by Blendle (described in Sec.~\ref{sec:commercial_solutions}), which is to offer readers an easy refund option on purchased articles as a way to reward quality content.

Unfortunately there has been very little research done on mental models for micropayments systems to date. This is a widely neglected area that could use input from the fields of psychology, human-computer interaction (HCI), and behavioral economics. Understanding the psychology behind micropayments will not only improve system usability but also assist in crafting appropriate business models and successful deployment strategies.

\subsection{Business Models and Deployment}

There are important open research questions regarding pricing for a micropayments ecosystem. For example, is there perhaps a pricing threshold at which micropayments become viable? While there is currently no definite answer to this question, there may be some evidence for it. The Chicago Sun-Times launched a Bitcoin-based micropayment donation option for 24 hours in February 2014 and collected over 700 payments, ranging from a penny to over \$ 1000 \cite{morris2014bitcoin}. 63\% of these payments were for 25 cents, the apparent ``sweet spot''.

The question of how to optimally monetize digital goods and services is itself an open and active area of research. Novel business models are emerging for digital content beyond the traditional subscription and pay-as-you-go paradigms and interesting results are being reported. We present a brief overview on this topic in Appendix~\ref{sec:webpublishingmodelsprimer}. Readers interested in a more detailed discussion on this topic are referred to surveys \cite{bhattacharjee2011digital} \cite{lambrecht2014firms}.

Regarding mass deployment of micropayments system, there are certain desirable properties a system should have that would great help with success. For instance, a micropayments solution should be inter-operable across a wide range of merchants and integrate with existing payment infrastructures. Not only does this give consumers more opportunities to use the solution, but it will also open up new sources of revenue for merchants.

Standardization is an essential step in that direction, and that has been recognized by the W3C who have renewed their efforts in this direction with the recent launch of the Web Payments Working Group (described in Appendix.~\ref{sec:appendixstandardization}). An alternative proposal is to interconnect different micropayments systems using payments gateways, allowing conversion, collaboration and interoperability \cite{parhonyi2004collaborative}.

The current marketplace also requires that any solution should be supported on multiple platforms. This is for two reasons: the customer should not be restricted to only being able to make micropayments on a single platform only (this was a marked drawback of certain first-generation schemes like Millicent). Second, to cut down on roll-out costs and the network effects problem, Odlyzko suggests that it is a good strategy to piggyback micropayments onto an existing and widely-used infrastructure \cite{odlyzko2003case}. Mobile phones are an ideal candidate. Smartphone usage is high and phones now support considerable more resources (computing, memory, bandwidth). Currently there is also a tremendous opportunity here for merchants: a Gartner study \cite{gartner2015gartner} predicts that by 2017, mobile payments will make up to 50\% of e-commerce revenue in the United States. Goldman Sachs predicts that by 2018 mobile phone purchases will constitute 50\% of e-commerce globally \cite{bill2014mobile}.

This may help address the chicken-and-egg problem with micropayments systems. Users are more likely to trust and adopt new payment systems if they have a positive reputation \cite{abrazhevich2004electronic}, but merchants are reluctant to adopt new payment systems unless they are widely used. Piggybacking micropayment solutions onto the mobile phone infrastructure (or social media) may solve this problem.


\section{Conclusion}
\label{sec:conclusion}

In this paper we have undertaken a detailed survey of micropayments systems. We discuss their security properties and briefly document their development. Next we classify the multitude of research-based cryptographic and commercial systems as per their salient features and describe in detail the workings of representative solutions and highlight the intuition behind them. This is followed by a discussion of outstanding challenges for micropayment systems, important gaps in research, and relevant recommendations.

Our intention in this paper has been not just to provide a comprehensive technical resource, but also to highlight lessons from the past and articulate promising new directions. For this reason, we do not restrict our study to the cryptographic literature, but bring together and systematize critically important insights from a variety of fields impacting micropayments.

We hope our work has a positive impact on the future development of these systems.

\bibliographystyle{unsrt}

\begin{appendices}

\section{A Primer on Business Models for Web Publishing}
\label{sec:webpublishingmodelsprimer}

\textbf{Advertising:} The dominant model for Web publishing is along the lines of television and print media, i.e. by \textbf{incorporating advertisements} or ``selling eyeballs''. Advertisements constitute the primary revenue stream for the Web's most popular sites, such as Google, YouTube, and Facebook. Advertisements are incorporated in the content, along sidebars or before videos start, and the customer does not pay anything from her own pocket. This open access is appealing to customers. However, it is technically incorrect to conceive of these services as `free'. Advertising costs are passed down to the customer in the prices of the advertised goods.

This model has several positives: highly customized advertisements can be delivered very easily to consumers. Consumer behavior can also be easily tracked, providing a rich source of information for merchants and businesses. Furthermore, there are minimal mental transaction costs for the customer.
However, there are also a number of negatives. The tracking of consumer behaviour often infringes on privacy, sometimes in noticable ways when consumers are delivered targeted advertisments that may leak information to onlookers about their private behaviour. Video advertisments may waste the consumer's time, and use an unacceptable amount of bandwidth for some mobile users. Increasing numbers of consumers are using ad-blocking software, which is a serious threat to businesses using this model.

\textbf{Paywalls:} The second prominent model for Web publishing is via \textbf{paywalls}, where users are directly charged to access digital content. Paywalls can take on several forms. The most popular model is via \textbf{subscription}. This model is used by well-known brands such as Spotify, NetFlix, which charge members monthly fees to access their media catalogue.
Then there are soft paywalls (also referred to as Freemium services), which offer differentiated service. Users can access certain basic content for free, but to access premium or customized content or service requires a subscription or a payment. Free content may be actual content or a preview or free trial. Premium content may include complete articles or quicker access (as in the case of stock quotes) or even customized content (such as digital newspapers for the IPad). This is the route taken by brands such as the Wall Street Journal, the New York Times, and Amazon's Prime service. Some firms which are primarily rely on the advertisement-based model may offer an ad-free service for subscribers. YouTube has launched such a paywall for consumers in the US \cite{CNet2016} at \$9.99 a month, and intends to extend this to the rest of the world at a later date. Google Contribute is a similar offering.
A third type of paywall is the pay-as-you-go paywall in which a charge is levied for each article read.

Paywalls have yielded mixed results thus far and they are recognized as being a difficult strategy to implement. The paid-subscription model has notably failed in several cases, such as with the New York Times' Times Select \cite{reagan2010five}, Time \cite{halliday2010time}, and The Sun.
It has proved successful in other instances, such as when the New York Times retried it in 2011, and with ESPN and The Spectator. The reasons for success and failure are still being debated, but quality of content appears to be a strong factor.

\section{Standardization}
\label{sec:appendixstandardization}

In the mid 1990s, the World Wide Web Consortium (W3C) \cite{w3c} (the predominant international standards organization for the Web, consisting of technology firms, merchants, research laboratories and standards bodies, including support from the European Commission and DARPA) launched a Micropayment Markup Working Group (MPM-WG).  This working group developed a Micropayment Transfer Protocol (MPTP) \cite{hallam1995micro} to handle money transfers online in a secure manner, and a language to embed micropayment initiating instructions in web pages, Common Mark-up for Micropayment per-fee-links \cite{world2002common}. None of these contributions became full standards, but they are available in the public domain, and some of these ideas were implemented in certain micropayment solutions, such as the NewGenPay micropayments system. The Working Group ceased activities in 2001.

The PayCircle consortium, founding members of which included CSG Systems, Hewlett-Packard, Oracle, Siemens, and Sun Microsystems, commenced work on developing standards for mobile payments and micropayments in 2002 \cite{paycircle2005}. PayCircle developed open application interfaces (APIs) to enable software developers to build universal payment-enabled applications for mobile business which interoperate with payment service providers such as telecom operators and banks. A public draft was submitted for consideration to the Open Mobile Alliance (OMA). PayCircle concluded operations in 2005.

Secure Mobile Payment Service (SEMOPS) was an EU funded project formed in 2002 by banks, technology companies, and research institutions, including names such as Motorolla, Deloitte and Millenium Bank. It aimed to develop universal electronic payment solutions for peer-to-peer payments, mobile and Internet payments, real-time payment transfers between accounts, and micropayments. The project lasted two years and their contributions are described in \cite{vilmos2003semops} \cite{karnouskos2004semops}. A follow-up project ran from 2007-08 with a focus on launching mobile payment services in some European countries.

In October, 2015, the W3C launched the Web Payments Working Group \cite{w3c2015w3c}, to develop standards which ``will support a wide array of existing and future payment methods, including debit, credit, mobile payment systems, escrow, and Bitcoin and other distributed ledger technologies.'' Micropayments were one issue they discussed at TPAC 2016 \cite{TPAC2016Minutes}.

\end{appendices}

\end{document}